%% file: glpapppl1.tex
\documentstyle[12pt]{article}
\date{}
\input psfig.sty

\begin{document}

\title{Light Baryons in a Constituent Quark Model with Chiral Dynamics}

\author{L. Ya. Glozman$^1$, Z. Papp$^2$, W. Plessas$^1$}
\maketitle

\centerline{\it $^1$ Institute for Theoretical Physics, 
University of Graz,}
\centerline{\it Universit\"atsplatz 5, A-8010 Graz, Austria}

\centerline{\it $^2$ Institute of Nuclear Research of the Hungarian
Academy of Sciences,} 
\centerline{\it Bem t\'er 18/c, P.O. Box 51, H-4001 Debrecen, Hungary}

\setcounter{page} {0}
\vspace{1cm}

\centerline{\bf Abstract}
\vspace{0.5cm}

It is shown from rigorous three-body Faddeev calculations that the masses
of all 14 lowest states in the $N$ and $\Delta$ spectra can be described 
within a constituent quark model with a Goldstone-boson-exchange interaction
plus linear confinement  between the constituent quarks.\\

PACS numbers 12.39.-x, 12.38.Lg, 12. 40.Yx, 14.20.Gk

\bigskip
\bigskip

Submitted to Physics Letters B

\bigskip
\bigskip
\noindent
\small

Preprint UNIGRAZ-UTP 27-01-96\\


\newpage
{\baselineskip=0.7cm

Recently it has been suggested \cite{GLO1,GLO2} that beyond the scale of
spontaneous breaking of chiral symmetry ($SB\chi S$) light and 
strange baryons can be 
viewed as systems of three constituent quarks which  interact
by the exchange of Goldstone bosons (pseudoscalar mesons) and are subject
to confinement. This concept is motivated by the idea that
instead of operating with the original QCD degrees of freedom,
i.e. current quarks and gluon fields, it is appropriate,
in low-energy QCD, to deal with constituent quarks and chiral boson fields. 
This is  in close analogy
to solid-state physics where instead of considering original degrees
of freedom (light electrons, a lattice of ions and the electromagnetic
field) one works with "heavy" electrons with dynamical mass,
phonons and their interactions.\\

In QCD the approximate $SU(3)_L \times SU(3)_R$ chiral symmetry in the (u,d,s)
sector is realized in the hidden Nambu-Goldstone mode at low temperature and
density, i.e. is spontaneously broken up to $SU(3)_V$. This is manifested
in the existence of the octet of low-mass pseudoscalar mesons representing
approximate Goldstone bosons. The flavor singlet $\eta'$ gets decoupled from 
the original $U(3)$ nonet due to instanton fluctuations in the QCD
vacuum \cite{THO}. With respect to the fermion fields, the effect of
$SB\chi S$ is that the valence current quarks acquire a dynamical 
mass \cite{WEIN,MAG}. As it was suggested within
a microscopical approach \cite{DIP}, this occurs when quarks propagate 
through instantons in the QCD vacuum \cite{SHU}. Thus in low-energy 
QCD one may rely on a chiral Lagrangian with dynamical fermion
fields (constituent quarks) and nine chiral fields. Of course, the full
effective chiral Lagrangian is not known. However, any possible (ps or pv)
coupling of the constituent quarks and chiral meson fields will imply a 
Goldstone-boson exchange (GBE) interaction  between the constituent quarks
when the chiral meson fields are integrated out of the Fock space.\\

Already
in refs. \cite{GLO1,GLO2} it has been indicated that this GBE interaction
will be capable of describing all features of the light and strange
baryon spectra. In particular, it has been shown how the spin-flavor
symmetry of the GBE can resolve some long-standing problems 
in baryon spectroscopy.
By taking into account the GBE within  first-order
perturbation theory one could  explain (i) the correct level ordering
of the positive- and negative-parity states in light and strange baryon
spectra as well as the approximate splittings in those spectra and (ii)
the branching ratios of $N^* \rightarrow N\eta$,
$\Lambda^* \rightarrow \Lambda\eta$ and  $\Sigma^* \rightarrow \Sigma\eta$
decays \cite{GLO3}.\\

In the present work we report results for the light baryon spectra
obtained from rigorous three-body Faddeev calculations. Besides  the
confinement potential, which is here taken in linear form, the GBE 
interaction between the constituent quarks is taken into account
to all orders. Our results  further
support the adequacy of the GBE for  baryon spectroscopy.
In particular, we shall demonstrate the description of  all 14 
lowest states in the $N$ and $\Delta$ spectra in agreement with 
phenomenology.\\

For the GBE the spin-spin component of the interaction
between the constituent quarks $i$ and $j$ consists of the octet and singlet
exchange components:

$$V_\chi^{octet}(\vec r_{ij})  =
\left\{\sum_{a=1}^3 V_{\pi}(\vec r_{ij}) \lambda_i^a \lambda_j^a
+\sum_{a=4}^7 V_K(\vec r_{ij}) \lambda_i^a \lambda_j^a
+V_{\eta}(\vec r_{ij}) \lambda_i^8 \lambda_j^8\right\}
\vec\sigma_i\cdot\vec\sigma_j, \eqno(1)$$

$$V_\chi^{singlet}(\vec r_{ij}) = \frac {2}{3}\vec\sigma_i\cdot\vec\sigma_j 
V_{\eta'}(\vec r_{ij}).\eqno(2)$$

\noindent
Here $\vec\sigma$ and $\vec \lambda$ are the quark spin and flavor matrices,
respectively. In the simplest case, when the boson field satisfies 
the linear Klein-Gordon equation, one has the following spatial dependence 
for the meson-exchange potentials in (1) and (2): 

$$V_\gamma (\vec r_{ij})= \frac{g_\gamma^2}{4\pi}\frac{1}{3}\frac{1}{4m_im_j}
\{\mu_\gamma^2\frac{e^{-\mu_\gamma r_{ij}}}{ r_{ij}}-4\pi\delta (\vec r_{ij})\}
,\eqno(3)$$

$$ (\gamma= \pi, K, \eta, \eta'), $$

\noindent
with quark and meson masses $m_i$ and $\mu_\gamma$, respectively.
We refer to paper \cite{GLO2} for a detailed
description of the properties of the octet-exchange interaction (1) within
and beyond the $SU(3)$ limit.\\

Eq. (3) contains both the traditional long-range 
Yukawa potential as well as a 
$\delta$-function term. It is the latter  that is of crucial importance 
for baryon physics.
It is strictly  valid  only for pointlike particles.
It must be smeared out, however, as the constituent
quarks and pseudoscalar mesons have in fact finite size.
Further it is quite natural to assume that at distances
$r \ll r_0$, where $r_0$ can be related to the
constituent quark and pseudoscalar meson sizes, there is no chiral 
boson-exchange interaction as this is the region of perturbative QCD with 
original QCD degrees of freedom. The interactions at these very
short distances are not
essential for the low-energy properties of baryons. Consequently
we use a two-parameter
representation for the $\delta$-function  term in (3)

$$4\pi \delta(\vec r_{ij}) \Rightarrow \frac {4}{\sqrt {\pi}}
\alpha^3 \exp(-\alpha^2(r-r_0)^2). \eqno(4)$$

\noindent
Obviously this substitution is exact in the limit 
$r_0 \rightarrow 0, ~ \alpha \rightarrow \infty$.
Following the arguments above one should also 
cut off the Yukawa part 
of the GBE for $r < r_0$. In order  to avoid any cut-off
parameter we use a step-function cut-off at $r = r_0$.
It may be introduced by the analytically smooth representation
$\left[ 1-(1+\exp(\beta(r-r_0)))^{-1}\right]^5$
with $\beta \rightarrow \infty$. We have tested that increasing
$\beta$ beyond the actually employed value of $\beta = 20$ fm$^{-1}$
does not change the results noticeably.
For the meson masses in (3) we take their physical
values, $\mu_\pi = 139$ MeV, $\mu_\eta = 547$ MeV, $\mu_{\eta'} = 958$ MeV,
and for the u and d constituent quarks we use $m=340$ MeV, like it is
suggested from nucleon magnetic moments.\\

The $\pi q$ coupling constant can be extracted from 
the phenomenological pion-nucleon 
coupling \cite{GLO2} as $\frac{g_8^2}{4\pi} =0.67$. For simplicity
(and to avoid any additional free parameter) the same coupling constant
is assumed for the  coupling between the $\eta$- meson and the constituent 
quark. 
This is  exactly in the spirit of unbroken
$SU(3)_F$ symmetry.  For the flavor-singlet $\eta'$, however,
we must take a different coupling
$\frac{g_0^2}{4\pi}$, as the $U(3)_V$ symmetry is known 
to be very strongly broken
to $SU(3)_V \times U(1)_V$. This fact is best illustrated by the failure of 
the Gell-Mann-Oakes-Renner relations \cite{GNOR} for
the flavor singlet \cite{WE}. Lacking a
phenomenological value, we treat $g_0^2/4\pi$ as a free parameter.\\

In the present calculation we neglect tensor meson-exchange forces. We expect
their
role to be of minor importance for the main features of baryon
spectra \cite{GLO2}  (mainly due to
the absence of the strong $\delta$ - function part in this case).
We emphasize that the primary aim in this paper is to demonstrate
the feasibility and advantages of the GBE model in baryon spectroscopy.
For this purpose the dynamical ingredients of (1) and (2) are already
sufficient.\\

Our full interquark potential is thus given by

$$V(\vec r_{ij})= V_\chi^{octet}(\vec r_{ij}) + 
V_\chi^{singlet}(\vec r_{ij}) + Cr_{ij}. \eqno(5)$$

\noindent
While all masses and the octet coupling constant are predetermined,
we treated
$r_0, ~\alpha,~ (g_0/g_8)^2,$ and $C$ as free parameters and determined
their values to be:

$$r_0 = 0.43 fm, ~\alpha = 2.91 fm^{-1},~ 
(g_0/g_8)^2= 1.8,~ C= 0.474 fm^{-2}.$$

Notice that we do not need any constant $V_0$, which is usually
added to the confining potential. In fact only  {\it four} free
parameters  suffice to describe all 14 lowest states of the $N$ and $\Delta$
spectra, including the absolute value of the nucleon (ground state).\\

The quark-quark potential (5) constitutes the dynamical input into our 
3-body Faddeev calculations of the baryon spectra. The Faddeev equations
were solved along the new method of ref. \cite{PP}, 
designed for an efficient solution of any 3-body bound state problem.
It has already been successfully
employed  in atomic and nuclear problems before.
We have carefully checked the accuracy of the results for all baryon levels,
in particular we have ensured perfect convergence with respect to
subsystem angular-momentum states included \footnote
{ In the meantime the results have been cross-checked with the help of 
the stochastic variational method \cite{VARGA}, and excellent agreement 
was found.}.
\\

We show our results  in Table 1 and Fig. 1
for the  parametrization of the qq
interaction given above.
It is well seen that the whole set of lowest $N$ and $\Delta$
states is quite correctly reproduced. In the most 
unfavourable cases deviations  from the experimental 
values do not exceed 3\% ! In addition all level orderings are 
correct. In particular, the positive-parity state $N(1440)$ (Roper resonance)
lies {\it below} the pair of negative-parity states $N(1535)$ - $N(1520)$. The
same is true in the $\Delta$ spectrum with $\Delta(1600)$ and the pair
$\Delta(1620)$ - $\Delta(1700)$. Thus a long-standing problem of baryon
spectroscopy is now definitely resolved. 
We emphasize again that
the qq potential (5) is able to predict also the absolute value of the nucleon
mass. In previous models an arbitrary constant was usually needed to achieve 
the correct value of 939 MeV.\\

At the present stage of our investigation of the baryon spectra with the
GBE interaction we have left out the tensor forces. Therefore the 
fine-structure splittings in the $LS$-multiplets are not yet introduced.
However, it is clear from the observed smallness of these splittings
and from the arguments given above, that the tensor component of the GBE
can play only a minor role. Here we also note that the Yukawa part of
the interaction in  (3) is only of secondary importance. In fact, the pattern
of Fig. 1 could also be described with the $\delta$-part  (4)
alone (and a  slightly modified set of parameters).\\

It is instructive to learn how the GBE affects the energy levels when
it is switched on and its strength (coupling constant) is gradually
increased (Fig. 2). Starting out from the case with confinement only, 
one observes that the degeneracy of states is removed and the 
inversion of the ordering of positive- and negative-parity states
is achieved, both in the $N$ and $\Delta$ excitations. The reason for 
this behaviour lies in the flavor dependence of the GBE (see a more
detailed discussion in ref. \cite{GLO2}). From Fig. 2 also the
crucial importance of the chiral interaction $V_\chi$
becomes evident. Notice that
the strength of our confinement, $C$= 0.474 fm$^{-2}$, is rather small and
the confining interaction contributes much less to the splittings than
the GBE.\\

We have also calculated the nucleon and delta  root mean square "matter" radii,
$\sqrt<r_N^2> = 0.465$ fm and $\sqrt<r_\Delta^2> = 0.54$ fm. 
For the nucleon one may view this result vis-\`a-vis the axial r.m.s.
radius $\sqrt<r_{axial}^2> = 0.68(2)$ fm and also the proton charge
r.m.s. radius $\sqrt<r_p^2> = 0.862(12)$ fm. It is clear that our result
must be smaller, as both of these phenomenological 
values include (additive and non-additive)
effects from the finite size of the constituent quarks and from meson-exchange
currents. It is convincing to observe, however, 
that we obtain $r_N < r_\Delta$,
in line with expectations.\\

A natural question to be addressed at this point is the role of 
one-gluon-exchange (OGE) between the constituent quarks \cite{DERU}.
Evidently, OGE alone cannot, in principle,  describe the $N$ and $\Delta$
spectra correctly \cite{GLO4}. One may ask, however, if there is an
additional contribution of OGE beyond GBE. Such a situation was investigated
in a very recent work \cite{DFM}. There a OGE with a 
relatively strong coupling
constant of $\alpha_s = 0.7$ was assumed in addition to GBE. A comparison
to the results in ref. \cite{DFM} reveals that our model with GBE alone
is capable of describing the $N$ and $\Delta$ spectra in a much better
and more comprehensive manner. While the authors of ref. \cite{DFM}
consider only 5 splittings, we are able to reproduce the masses of the
14 lowest $N$ and $\Delta$ states. Clearly, we might include
small effects from OGE (i.e. with a rather weak, but realistic, coupling
$\alpha_s = 0.1 - 0.2$, say) but  OGE would  then be negligible relative to
GBE.
Any sizeable coupling
$\alpha_s  > 0.2 -0.3$ would spoil the
spectra, however. This  is most dramatically seen in the ordering of the 
positive-parity state
$\Delta(1600)$ and the negative-parity $\Delta(1620) - \Delta(1700)$ pair.
The color-magnetic interaction shifts the $\Delta(1600)$ strongly up
but not down with respect to  $\Delta(1620) - \Delta(1700)$ \cite{GLO4}. 
In addition,
the OGE with large $\alpha_s$  would imply 
strong spin-orbit forces while there are practically
no spin-orbit splittings in the $N$ and $\Delta$ spectra. Note also that the
effect of $\eta'$ exchange cannot be generated by the OGE,
as the color-magnetic contribution  from OGE has an opposite sign.\\

For the $\eta'$-quark coupling $\frac{g_0^2}{4\pi}$ we have a value
that is  bigger
than for $\pi$-quark coupling. This is quite natural to
expect as the $\eta'$ mass is  twice larger than the $U(3)$ current
algebra prediction (due to the U(1) anomaly). It is comforting that our
value for $(g_0/g_8)^2$ is within the limits deduced 
in refs. \cite{CHENG1, CHENG2} from the
Gottfried sum-rule violation, the $\bar u/ \bar d$ ratio, as well as
the flavor and spin contents of the nucleon. The main idea of that work and a 
previous
paper \cite{QUIGG} is that the  content of the quark sea in the nucleon
is governed by the coupling between valence quarks
and goldstone bosons.\\
 
Our concept of GBE between constituent quarks may also be viewed
in the light of some recent evidences from lattice QCD.
It was shown  that one can obtain a
qualitatively correct splitting between $N$ and $\Delta$ already within
quenched approximation \cite{WEINGART}.
However, it was not clear from this result what where the
physical reasons for this splitting, OGE, instantons, or something else.
To clarify this question, the $N$ - $\Delta$ splitting 
has recently been measured in the quenched and a further,
so-called valence, approximation \cite{LIU}. 
While the standard quenched approximation contains part of
antiquark effects related to $Z$ graphs formed of valence quark lines, 
in the valence approximation   the quarks are
limited to propagate only forward in time (i.e.  $Z$ graphs and
corresponding $q\bar q$ pairs do not enter). 
The gluon exchange and all other possible
gluon configurations, including instantons, are exactly the same in both
approximations. The striking result is that the
$N-\Delta$ splitting is observed only
in the quenched approximation but not in the valence approximation,
in which  the $N$ and the $\Delta$ levels are degenerate within error bars.
Consequently the $N-\Delta$ splitting must receive a considerable
contribution from
$q\bar q$ excitations but
not from the gluon exchanges or instanton-induced interactions between
the constituent quarks.\\

Summarizing, in this paper we have presented first results
for baryon spectra obtained from the GBE model within rigorous 3-particle
Faddeev calculations. We relied on an interaction between the
constituent quarks stemming from the exchange of octet and singlet
pseudoscalar mesons. A simple parametrization of its 
spin-spin component was enough to achieve the absolute values of the masses of
the 14 lowest states in the $N$ and $\Delta$ spectra. The efficiency
of the model is evidently due to the spin-flavor dependence introduced
by GBE. It leads to an immediate resolution of hitherto persisting
problems in baryon spectroscopy. Certainly, the GBE model needs to be 
extended and refined further. Given the indications
of refs. \cite{GLO1, GLO2} we expect to be able 
to explain also the strange baryon
spectra. For the fine-structure splittings of the LS multiplets and detailed
properties of baryon structure, obviously, also the tensor 
forces have to be included.
Corresponding work in both of these directions is presently under way.\\

The authors are grateful to K. Varga for checking the results with his 
stochastic variational method. L.Ya.G. is thankful to D.O. Riska for many
stimulating discussions and to G.E. Miller
for communicating the results of ref. \cite{DFM}. 
This work was partially supported by the 
Austrian-Hungarian Scientific-Technical Cooperation within project
A23 (Austria) resp. 17/95 (Hungary). The work of Z.P. was also supported
by OTKA grant T17298.

\
\newpage
\centerline{\bf Table 1}

Masses of the 14 lowest states for non-strange
three-quark systems with total orbital angular momentum
L, spin S, and isospin T.
\begin{center}
\begin{tabular}{|llll|} \hline
&&&\\
$^{2T+1~ 2S+1}L $ & Parity & Theoretical & Corresponding phenomenological\\ 
& & Mass in MeV & LS multiplet\\
&&&\\ \hline
&&&\\
$ ^{22}S$ & + & 939 & $N$\\
&&&\\
$ ^{44}S$ & + & 1232 & $\Delta$\\
&&&\\
$ ^{22}S$ & + & 1493 & $N(1440)$\\
&&&\\
$ ^{44}S$ & + & 1635& $\Delta(1600)$\\
&&&\\
$ ^{22}P$ & -- & 1539 & $N(1535)-N(1520)$\\
&&&\\
$ ^{42}P$ & -- & 1667 & $\Delta(1620)-\Delta(1700)$\\
&&&\\
$ ^{24}P$ & -- & 1640 & $N(1650)-N(1700)-N(1675)$\\
&&&\\
$ ^{22}D$ & + & 1635 & $N(1680)-N(1720)$\\
&&&\\
$ ^{22}S$ & + & 1690 & $N(1710)$\\
&&&\\ \hline
\end{tabular}
\end{center}

\newpage

\begin{center}
{\bf Figure Captions}
\end{center}

\noindent
Fig.1 

\bigskip
Energy levels for the 14 lowest non-strange baryons with total
angular momentum and parity $J^P$. The shadowed boxes
represent experimental
uncertanties.

\bigskip
\bigskip

\noindent
Fig.2

\bigskip
Level shifts of some lowest baryons as a function
of the  strength of the GBE. Solid and dashed
lines correspond to positive- and negative-parity states, respectively.
}
\newpage
\begin{figure}
\input baryon.tex
\vspace{2cm}
\caption{$$}
\end{figure}
\newpage

\begin{figure}

$$\hfill \psfig{file=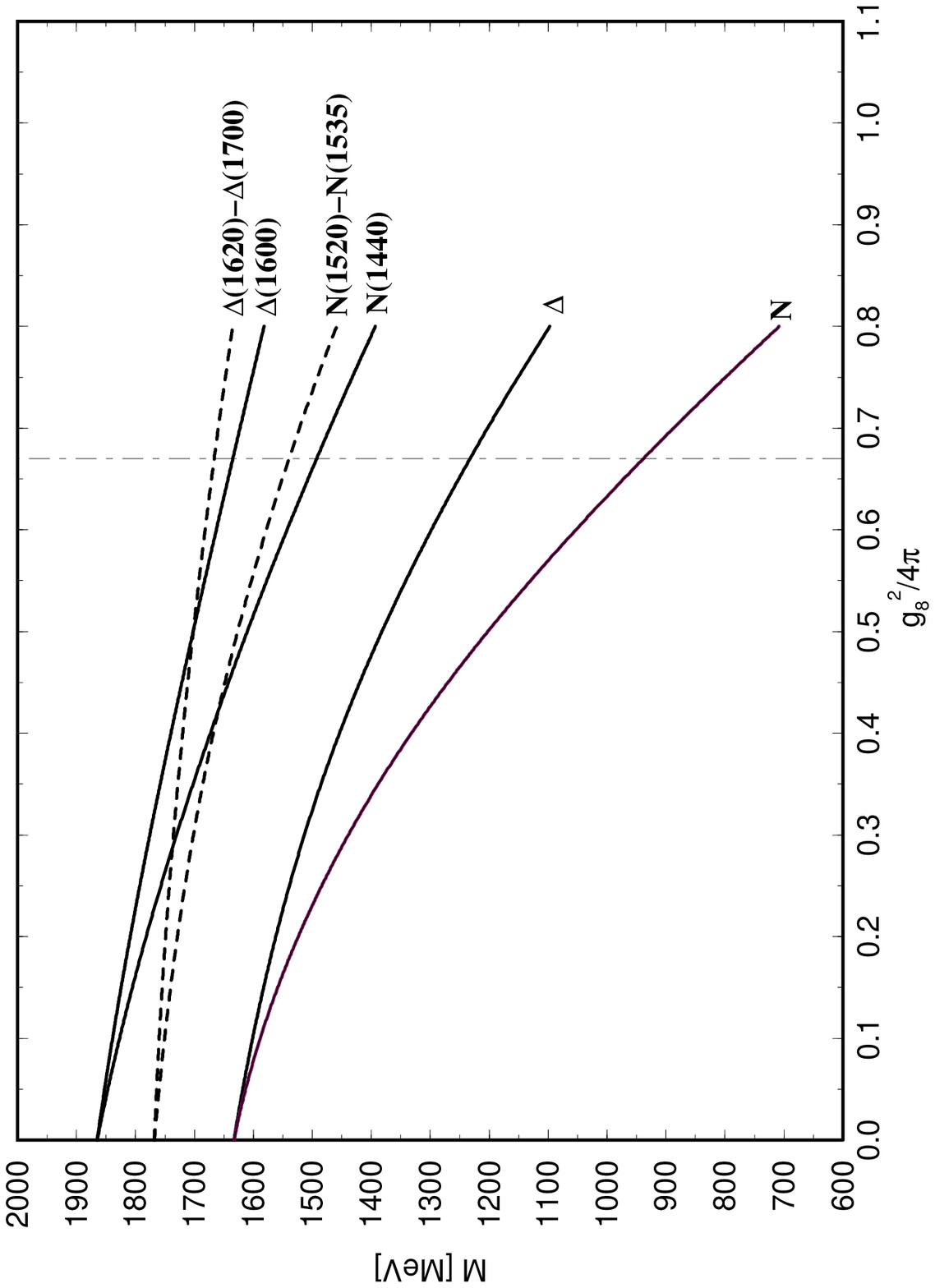} \hfill $$
\caption{$ $}
\end{figure}

\end{document}

%% file: baryon.tex
%
%
%
%
\input pictex.sty

%
%

%
%
\beginpicture

%
%
%
%
%
%
%
%
%

\setcoordinatesystem units <1mm,0.1mm>

%
%
\setplotarea x from 0 to 110, y from 900 to 1900

%
%
\axis left label {\stack{M,[MeV]}} ticks withvalues 
$900$ 
$1000$ 
$1100$ 
$1200$ 
$1300$ 
$1400$ 
$1500$ 
$1600$ 
$1700$ 
$1800$
 / 
at 900 1000 1100 1200 1300 1400 1500 1600 1700 1800 / /
%
{\Large
\axis bottom label {\Huge $\qquad N \qquad \qquad \qquad \quad \Delta$} ticks withvalues 
${\frac{1}{2}}^+$ 
${\frac{1}{2}}^-$
${\frac{3}{2}}^+$ 
${\frac{3}{2}}^-$ 
${\frac{5}{2}}^+$ 
${\frac{5}{2}}^-$ 
${\frac{1}{2}}^-$
${\frac{3}{2}}^+$ 
${\frac{3}{2}}^-$
/ 
at 10 20 30 40 50 60 80 90 100 / /
}
%
\shaderectangleson
\setshadegrid span <0.6mm>

%
%
%
\putrectangle corners at  7 1430 and  13 1470
\putrectangle corners at  7 1680 and  13 1740
\putrectangle corners at 17 1520 and  23 1555
\putrectangle corners at 17 1640 and  23 1680
\putrectangle corners at 27 1650 and  33 1750
\putrectangle corners at 37 1515 and  43 1530
\putrectangle corners at 37 1650 and  43 1750
\putrectangle corners at 47 1675 and  53 1690
\putrectangle corners at 57 1670 and  63 1685
\putrectangle corners at 77 1615 and  83 1675
\putrectangle corners at 87 1230 and  93 1235
\putrectangle corners at 87 1550 and  93 1700
\putrectangle corners at 97 1670 and 103 1770

\shaderectanglesoff

%
\linethickness0.3mm

%
\putrule from  7  940 to  13  940
\putrule from  7 1493 to  13 1493
\putrule from  7 1690 to  13 1690
\putrule from 17 1539 to  23 1539
\putrule from 17 1640 to  23 1640
\putrule from 27 1635 to  33 1635
\putrule from 37 1539 to  43 1539
\putrule from 37 1640 to  43 1640
\putrule from 47 1635 to  53 1635
\putrule from 57 1640 to  63 1640
\putrule from 77 1667 to  83 1667
\putrule from 87 1232 to  93 1232
\putrule from 87 1635 to  93 1635
\putrule from 97 1667 to 103 1667

%
\endpicture

%